\RequirePackage{fix-cm}
\documentclass[twocolumn]{svjour3}          
\smartqed  
\usepackage{graphicx}
\usepackage{epstopdf}
\usepackage{hyperref}
\usepackage{siunitx}   
\usepackage{booktabs} 
\usepackage{tabu}
\usepackage{color}
\begin{document}
 \newcommand{\vect}[1]{\mbox{{$\mathbf{#1}$}}}
\newcommand{\micron}{{$\mu$m}}
\newcommand{\micronsp}{{$\mu$m }}
\newcommand{\III}{{II$'$}}
\newcommand{\IIIsp}{{II$'$ }}
\newcommand{\Frac}[2]
	 {\displaystyle{\frac{\displaystyle{#1}}{\displaystyle{#2}}}} 
\newcommand{\Sum}[2]
	 {\displaystyle{\sum_{#1}^{#2}}} 
 \newcommand{\typeI}{Type I }
  \newcommand{\typeII}{Type II }
    \newcommand{\typeIIp}{Type II$'$ }

\title{Layer dependent role of collagen recruitment during loading of the rat bladder wall
}


\author{Fangzhou Cheng       \and
        Anne M. Robertson \and Lori Birder \and F. Aura Kullmann \and Jack Hornsby \and Paul Watton \and Simon C. Watkins
}


\institute{F. Author \at
              first address \\
              Tel.: +123-45-678910\\
              Fax: +123-45-678910\\
              \email{fauthor@example.com}           
           \and
           S. Author \at
              second address
}

\date{Received: date / Accepted: date}


\maketitle

\begin{abstract}
In this  work,  we reevaluated long standing conjectures  as to the  source of the exceptionally  large compliance  of  the  bladder  wall.  Whereas, these  conjectures  were based  on indirect  measures  of loading  mechanisms,   in  this  work we take  advantage of advances  in bioimaging  to  directly  assess collagen fibers and  wall architecture during loading.  A custom  biaxial  mechanical  testing  system  compatible with  multiphoton microscopy  (MPM) was used to directly  measure  the  layer  dependent  collagen fiber recruitment in bladder  tissue from 9 male Fischer  rats  (4 adult  and  5 aged).  As for other  soft tissues,  the  bladder  loading curve was exponential  in shape  and  could be divided  into  toe, transition and high stress regimes. The relationship between collagen recruitment and loading curves were evaluated in the context of the inner bladder  wall (lamina  propria) and outer detrusor smooth muscle layer. The  large extensibility of the bladder was found to be possible due to folds in the  wall (rugae) that  provide a mechanism for low resistance  flattening without any discernible recruitment of collagen fibers throughout the toe regime.     For elastic bladders, as the loading extended into the transition regime, a gradual coordinated recruitment of collagen fibers between  the  lamina propria  and  detrusor smooth  muscle  layers was found.  A second important finding is that wall extensibility  could be  lost by premature recruitment of collagen in the outer wall  that cut short the toe region.
This work provides, for the first time, a mechanistic understanding of the role of 
collagen recruitment in determining bladder capacitance.
\end{abstract}


\section{Introduction}
\label{intro}
The human bladder is a highly compliant organ that can expand from an empty state to store 450  $\pm$ 64.4 ml of urine at relatively low pressures of 43.5 $\pm$ 7.3 cm of H$_2$O in young adults (18 yrs)  \cite{Kershen2002}. This is quite remarkable given the low extensibility of the collagen fibers that make up the bulk of the extracellular matrix in the bladder wall.   
 However, the elasticity of the bladder wall can decrease with age  and disease  \cite{Ameda1999,Kohan2000}.
This increased stiffness not only diminishes the bladder capacity but also results in increased filling pressures that are transmitted to the kidney and can cause recurrent infection, abdominal pain and, in severe cases, reflux nephropathy or chronic kidney failure \cite{Andersson2011,Zeidel2016}.
These changes are also believed to contribute to incontinence, which is associated with large annual costs (estimated at nearly \$20 billion in 2000), \cite{Gorina2014,Hu2004}.  Hence, it is of great importance to understand the mechanical source of the bladder extensibilty and how this functional property is changed during aging and disease.
\par
Structurally, the bladder wall is a layered composite with passive components (elastin and collagen fibers) as well as active components (smooth muscle cells).  The components are distributed  
in  laminated structures within the bladder wall that are typically grouped into three layers, Fig. 1. Moving from the lumen, these are the mucosa, the muscularis propria (herein referred to as the detrusor smooth muscle layer) and the adventitia. The mucosa is composed of an urothelium, a basement membrane and a lamina propria (LP).    The LP  contains a densely packed, interwoven network of collagen I and III fibers \cite{Ewalt1992,Chang1999}.  
The detrusor smooth muscle layer (DSM) is a composite of smooth muscle bundles intermixed with type III collagen fibers and elastin fibers.  The collagen fibers are believed to interconnect the smooth cells within the bundles  \cite{Ewalt1992} and are less densely packed than in the lamina propria layer. The outer surface of the bladder is formed of  loose connective tissue commonly termed the adventitia.
\begin{figure}
\includegraphics[width=3.35in]{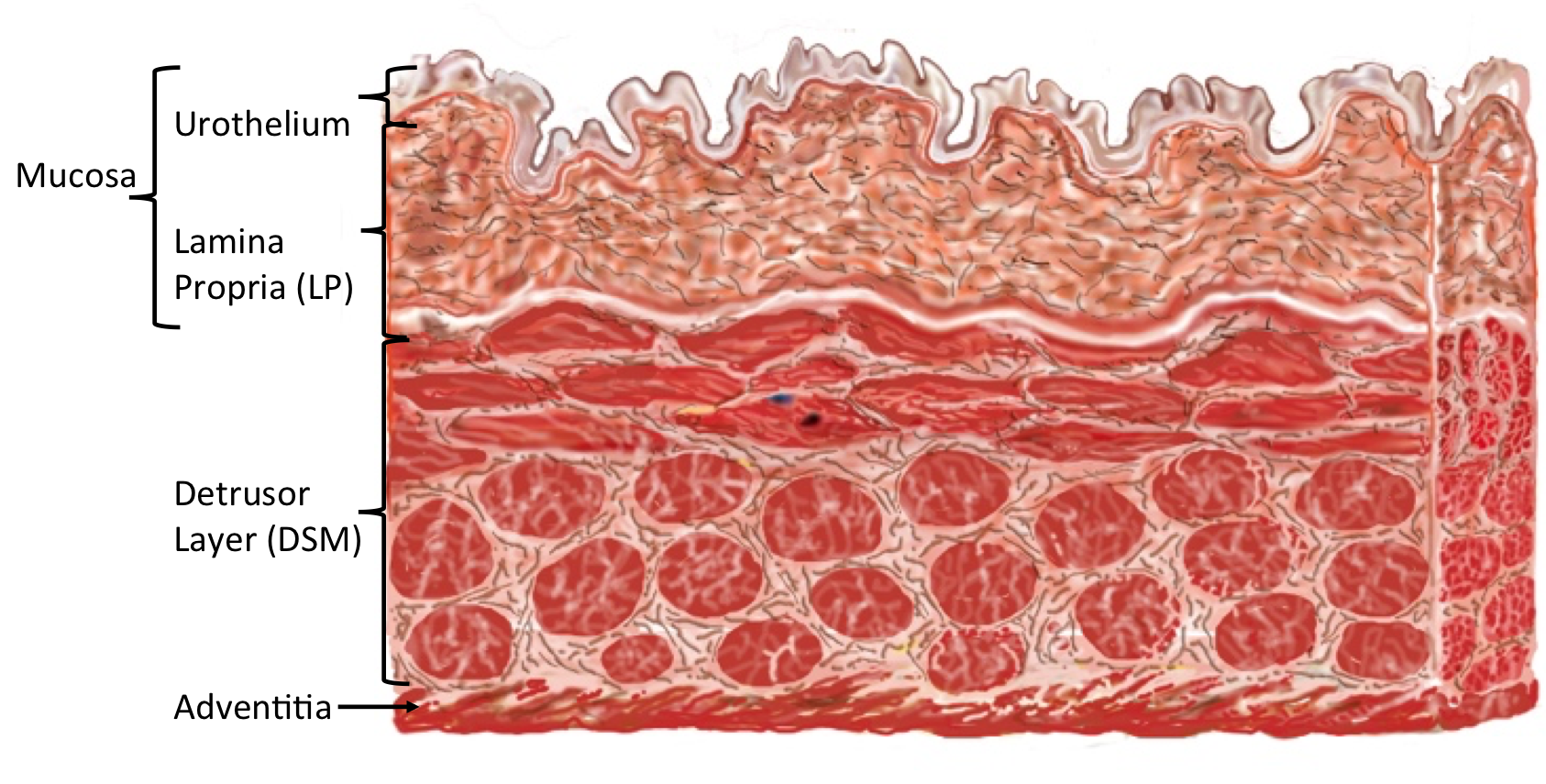}
\caption{Schematic of a cross section of the bladder wall showing the three major layers (mucosa, detrusor smooth muscle layer (DSM) and adventitia).  The folds (rugae) of the mucosa layer involve both the urothelium and LP layers. Smooth muscle bundles are shown aligned in two layers within the DSM.}
\label{fig:1}       
\end{figure}
\par
Even though it is generally accepted that bladder compliance is of tremendous clinical importance, a small number of studies performed nearly two decades ago are the main source of  information on the physical source of bladder wall extensibility. 
In 1992, Ewalt et al. published what appears to be  the first conjecture as to the source of bladder  compliance,  attributing  bladder extensibility to folds  in the LP collagen fibers, \cite{Ewalt1992}.  They conjectured the wavy LP fibers have little resistance to straightening, enabling the bladder to undergo large changes in volume with  small increases in pressure. They attributed the bladder filling capacity entirely to the  LP layer,  and in particular to the volume at which the LP collagen fibers are engaged.    
\par
Chang et al. explored  the conjecture of Ewalt et al.  by analyzing the collagen fibers in  bladders filled to  0\%, 25\%, 50\%, 75\% and 100\% of their total capacity  \cite{Chang1999}.  They similarly attributed bladder capacity  entirely to the lamina propria and made a second conjecture that the mechanical role of the detrusor layer is to simply serve as ``the limiting or girding layer to prevent over-distension of the bladder wall''.  This emphasis on the LP layer as the limiting factor for wall compliance was  primarily supported by a comparison of estimates of the maximum tension of the lamina propria and detrusor layer \cite{Susset1981} and their change in thickness during filling \cite{Chang1999}.   
These conjectures, while often referenced, have not been reevaluated,  (e.g. \cite{Aitken2009,Andersson2014}) with the exception of recent work on mouse bladder \cite{Hornsby2016}.   Hence, there is a need for further direct evidence  to prove or disprove these conjectures. 
\par
While there is a lack of direct measurements of the varied mechanical roles of the LP and DSM layers during bladder filling, changes to the extracellular matrix have been associated with altered mechanical function.  For example, Ewalt found that, in the non-compliant bladders of children ($\le 14$), the interfascicular space (between SMC bundles)  accumulates type III collagen fibers and elastin. He postulated these changes  prevent normal expansion of the detrusor layer and   thereby affect bladder compliance \cite{Ewalt1992}.
Non-compliant bladders have also been studied in the context of spinal cord injured and partial bladder obstruction \cite{Chen2013}. In these settings, non-compliant bladders were found to have increased density and volume of smooth muscle cells  \cite{Hald1998}, increased quantities of elastin, and decreased collagen fibers in the detrusor layer \cite{Nagatomi2004,Kim1991}. The ratio of collagen types I and III as well as  collagen fiber orientation were also found to be significantly different in the  detrusor layer of the non-compliant bladder relative to the compliant \cite{Kim1991,Chang1999,Ewalt1992}. These changes were speculated to directly or indirectly cause the loss of bladder compliance.  
\par  
In this work, we make use of advances in bioimaging coupled with mechanical testing to revisit these two conjectures.  In particular,
the purpose of the current study is to investigate the relationship between bladder wall structure and compliance during loading using recently developed imaging tools that enable concurrent imaging of collagen fibers and mechanical testing in intact, unfixed bladder samples \cite{Hill2012}.
We developed a planar biaxial system compatible with multiphoton imaging (MPM) that enables imaging of collagen fiber recruitment  during simulated bladder filling - approximated as a planar biaxial deformation. The structural and mechanical data were quantified and used to evaluate the mechanisms responsible for compliance and loss of compliance in rat bladders.  Attention was given to changing roles of the wall layers during loading from zero strain through supraphsyiological levels.
\par
An enhanced understanding of  the structural mechanisms responsible for  bladder compliance  will  guide the development of  novel medical treatments of bladder dysfunction, %
and provide design objectives for tissue engineered bladders aimed at mimicking the remarkable compliance of the bladder \cite{Pokrywczynska2015}. Furthermore, this study will identify the structural milieu of the intramural cells, knowledge that is essential for understanding the
 growth and remodeling process in health and disease  \cite{Aitken2009}. 

\section{Methods}
\label{sec:1}
\subsection{Tissue preparation}
\label{sec:1.1}
\begin{figure*}
\includegraphics[width=6.25in]{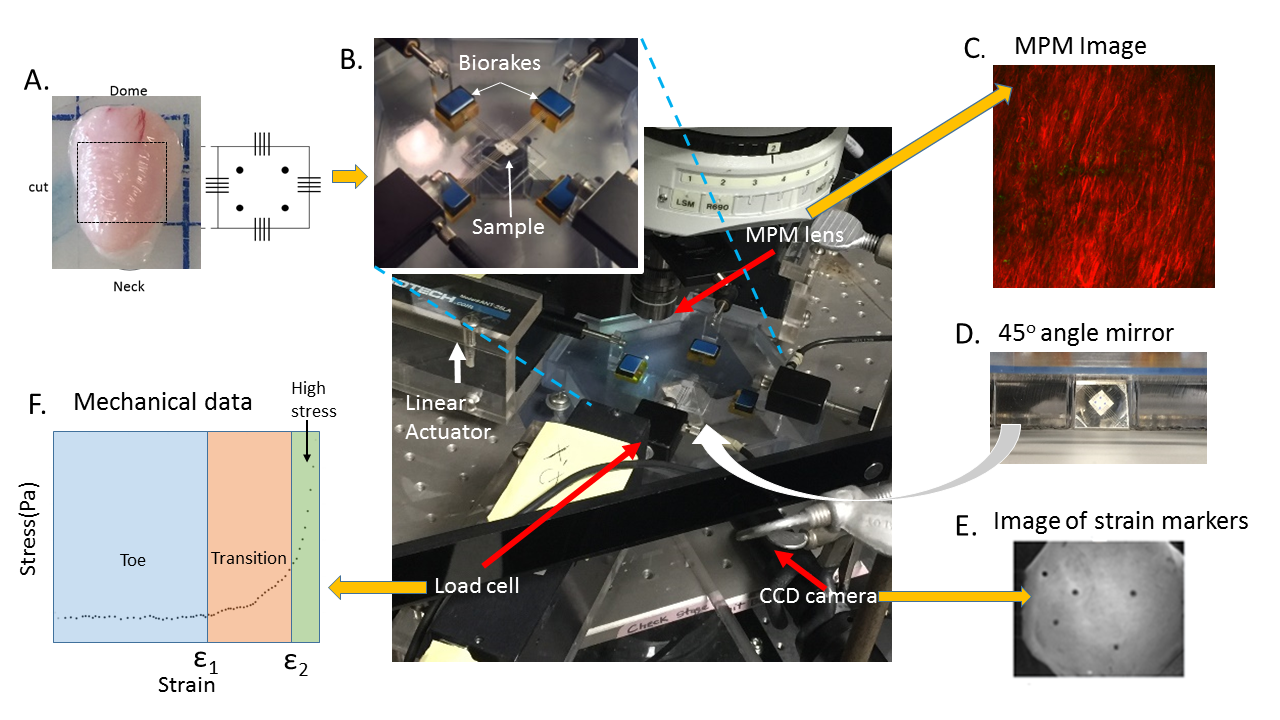}
\caption{Schematic of experimental system developed for mechanical testing and bioimaging of bladder samples.  Yellow arrows indicate  input (one) and   output (three) for the system.  (A) Bladder  and schematic of of square tissue sample  (input),  (B) Biaxial testing system under MPM lens with insert showing magnification of  tissue loaded on biorakes,  (C) Representative projected stack of images showing second harmonic generation  signal from MPM imaging (output), (D) Image from 45$^o$ mirror showing underside of loaded sample, (E) Image from CCD camera showing strain markers imaged using mirror system (output), (F) Representative loading stress strain curve showing toe, transition and high stress regimes of curve obtained using mechanical testing system (output).  }
\label{fig:2}       
\end{figure*}

Nine male Fischer rats (Species F344, adult and aged rats from National Institute of Aging of the NIH) were used in this study, separated by age with 4 adult rats (12 months) and 5 aged rats (21-24 month). The 
 unloaded meridional length  $(h_o)$ and diameter $(w_o)$ were measured in the explanted bladder. 
 The bladders were then cut open longitudinally and trimmed into square pieces with widths of 6mm $\pm$ 1mm such that the sides of the samples were aligned with the in situ longitudinal and circumferential directions, Fig. \ref{fig:2}A. To inhibit smooth muscle cell contraction, samples were immersed in Hanks Buffer Salt Solution (HBSS) containing, (in  mM) NaCl 138, KCl 5, KH2PO4 0.3, NaHCO3 4, MgCl2 1, HEPES 10, glucose 5.6, with pH 7.4, 310 mOsm/l) without calcium and with added EDTA (0.5mM). The voltage calcium channel blocker nifedipine (5uM; Sigma) and the SERCA pump inhibitor, thapsigargin (1uM; Tocris Biosciences), which prevents the reloading of intracellular calcium stores, were also added.
\subsection{Mechanical testing and constitutive modeling}
\label{sec:1.2}
Mechanical testing was performed using a custom biaxial system  specifically designed for testing  bladder tissue concurrent with imaging under a multiphoton microscope, Fig. \ref{fig:2}.  This design enabled imaging of collagen fibers in intact specimens without staining or fixation.  In the biaxial system, displacement can be independently controlled  by four actuators (Aerotech, Inc., linear actuator ANT-25LA) and  force measurements are performed using load cells on two of the actuators (Transducer techniques, nonrepeatibility 0.05\% of R.O., capacity 5 lbs), Fig.  \ref{fig:2}B. Tissue is mounted on the device using biorakes (World Precision Instruments, Inc.).  The biaxial system includes a CCD camera and a 45 degree offset mirror to enable imaging of strain markers from beneath the mounted tissue, Fig. \ref{fig:2}D,E.  This imaging system enables MPM imaging at prescribed biaxial strains.  
\par 
Prior to testing, the unloaded thickness $t_o$ of each sample was measured in 5 positions using a 0-150mm  digital caliper (Marathon watch company Ltd) and averaged. Fiducial strain markers (Basalt microspheres, 425-500 \micron, Whitehouse Scientific) were attached to the abluminal side of each sample for strain calculation.  
During testing, the square sample was first loaded lumen side up on the biorakes, Fig. \ref{fig:2}B.
Following Wognum et al., a tare-load was applied to the sample  after which it was preconditioned,  then unloaded,  then loaded to the tare-load, then mechanically tested \cite{Wognum2009}.
Five consecutive equibiaxial loading cycles to a stretch of 0.8 were used for preconditioning with a tare-load of 0.02N.  The lumen surface was imaged under MPM at stepwise increases in strains, Fig. \ref{fig:2}C,    (Section \ref{sec:MPM}).  To avoid tissue damage while obtaining a large range of strain, loading was stopped after collagen fibers  were visibly straightened (recruited).   Hence, an individual {maximum stretch} was identified for each sample. 
After lumen side imaging, 5 equibiaxial loading cycles to the maximum stretch were performed with the strain marker locations recorded by a CCD camera using the 45 degree mirror block beneath the sample (Fig. \ref{fig:2}D,E)  and used to obtain the loading curves. The sample was then flipped and the imaging was repeated from the abluminal, once again until collagen recruitment was observed. The components of the Green-Lagrange strains were calculated from the fiducial marker locations using a finite element interpolation method \cite{Sacks2000}.  Components of the Cauchy stress tensor were calculated using load measurements  with estimates of current cross sectional area obtained from the strain measurements under the approximation of isochoric motion.
\par
The  rat bladder was modeled as an incompressible,  hyperelastic, isotropic material  with  Cauchy stress tensor  
\begin{equation}
\vect{T}=-p\vect{I}+\mu \,  e^{\gamma(I_1-3)}  \vect{B}
\label{eq:exp}
\end{equation}
where  p is the Lagrange multiplier associated with incompressibility,
$I_1$ is the first invariant of the left Cauchy-Green deformation tensor  $\vect{B}$, $\mu_0$ is the shear modulus and material constant $\gamma$ controls the exponential dependence on $I_1$.  An exponential dependence  on stretch was  proposed by Fung \cite{Fung1967} and the form in (\ref{eq:exp}) is commonly used for  soft biological tissues.
 Data sets for the planar biaxial loading studies were combined for the longitudinal and circumferential directions and used to obtain the material constants in (\ref{eq:exp}).   
\subsection{Calculation of bladder compliance and capcity}
\label{sec:1.3new}
The International Continence Society (ICS) defines bladder compliance $C$  as 
\begin{equation}
C = \Frac{\Delta V}{\Delta P}
\label{eq:compliance}
\end{equation}
(given in units of ml per cm of H$_2$0) where $\Delta P$ is the change in intraluminal pressure between two different loading states and  $\Delta V$ is the corresponding change in volume \cite{Abrams2002}.
The initial loading state is chosen as the start of filling while the second and higher loading state is  either the cystometric capacity or the state immediately before any muscular bladder contractions have started (that would lead to substantial bladder leakage).  
Cystometric capacity is defined as the bladder volume when the patient has a normal desire to void, \cite{Abrams2002}. 
\par
In this work, consistent with the ICS definition for compliance,  we consider the two states to be the  zero pressure state  ($P_o$) and the  cystometric pressure   ($P_1$) set equal to  45 mmHg for rats,  \cite{Alexander1971}.  
The corresponding volumes are denoted $V_o$ and $V_1$, respectively.    The bladder capacity is simply $V_1$, namely, the amount of urine the bladder can hold at the cystometric pressure.  Since the compliance is calculated for a fixed cystometric pressure for a given population,  the compliance for this population  is simply the  bladder capacity multiplied by a constant.   
\par
In estimating the  volume at the zero pressure state ($V_o$), the bladder was approximated as an ellipsoid. 
An effective unloaded radius $r_o$ was then calculated from
\begin{equation}
r_o =  \Frac{1}{8} (w_o h_o)^{1/3}.
\end{equation}
The corresponding loaded volume was estimated from the analytic solution for pressure inflation of a spherical membrane \cite{Beatty87b} with bladder specific material constants obtained from the planar biaxial studies.  In particular,  the intraluminal  pressure for the exponential model as a function of stretch $\lambda$ 
\begin{equation}
P(\lambda) = \frac{2 t_o}{\lambda r_o} \, [1-\frac{1}{\lambda^6}]\mu e^{\gamma(I_1 - 3)}
\end{equation}
where  $\lambda$ is the ratio of the current  radius to $r_o$. 
\subsection{Assessment of regimes of the loading curve} \label{sec:1.3} Following the approach of Sacks \cite{Sacks2000}, the intramural stress in the bladder wall corresponding to filling pressure of 100KPa  estimated from an equilibrium balance for a sphere (Laplace's law).  Using the exponential fit, loading curves were extrapolated to this maximum load with corresponding maximum strain, $(\epsilon_{max})$
and strain, $\epsilon = (\lambda^2-1)/2$.  The average curve for each of the principal directions was calculated and divided into 3 regimes: toe regime, transition regime, and high stress regime as follows.  To determine the  strain defining the end of the toe regime, first a linear fit of the data starting from the origin was performed with an R$^2$ of 0.99.  The toe regime, or low stress regime was then defined by strains in the range [0,$\epsilon_1$),  where $\epsilon_1$ is the strain at which the loading data deviates from the  linear fit by 450 Pa.  The high stress regime was defined as $\epsilon  \in (\epsilon_2,\epsilon_{max}]$, Fig. 2F. Here $\epsilon_2$  was similarly defined relative to a linear fit to the data with an R$^2$ of 0.98, in this case, beginning with the maximum of the high stress data rather than the origin. The transition regime was defined as the intermediate regime,  $\epsilon \in [\epsilon_1, \epsilon_2$ ], Fig. \ref{fig:2}F.
\subsection{Multiphoton imaging}
\label{sec:1.3}
\label{sec:MPM}
Tissues were scanned with multiphoton imaging using a Z step size of 2 \micron, following methods in \cite{Hill2012}.  Briefly, a multiphoton microscope (Olympus FV1000 MPE) equipped with a Coherent Chameleon TiSapphire pulsed Laser  was used to image collagen fibers. An excitation wavelength of 800 nm and 1.12 NA 25 x MPE water immersion objective were used for all samples. Signals from second harmonic generation (SHG) were collected using a 400 nm emission filter with a  $\pm$50 nm spectral bin.   
\subsection{Evaluation of fiber recruitment from multiphoton images}
\begin{figure}[!ht]
\includegraphics[width=3.25in]{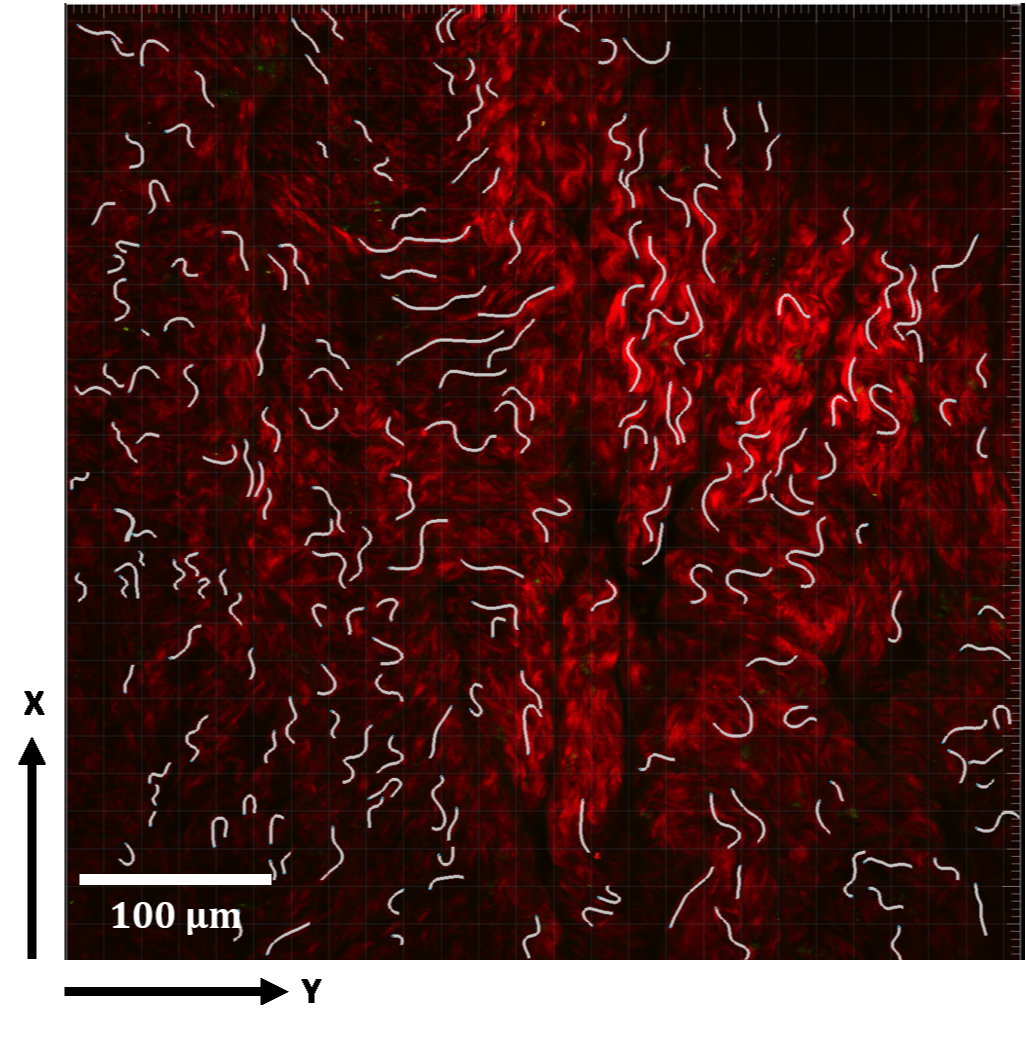}
\caption{Collagen fiber tracings shown in white in a 2D  projection of a representative 3D reconstructed model formed from 2D multiphoton stacks. MPM signal from collagen fibers seen in red. }
\label{fig:4}      
\end{figure}

Multiphoton image stacks were used to obtain 3D reconstructions of the collagen fiber architecture that were mapped to the corresponding point on the loading curve by matching the loading level. Collagen fibers could be clearly imaged up to depths of approximately 200 \micronsp in unloaded samples. Collagen fibers were  traced in 2D slices through the depth of the 3D reconstructed model (Filament function in Imaris, Bitplane, Switzerland), Fig. \ref{fig:4}, \cite{Hill2012}. 
\par
\begin{figure}[!ht]
\includegraphics[width=3.25in]{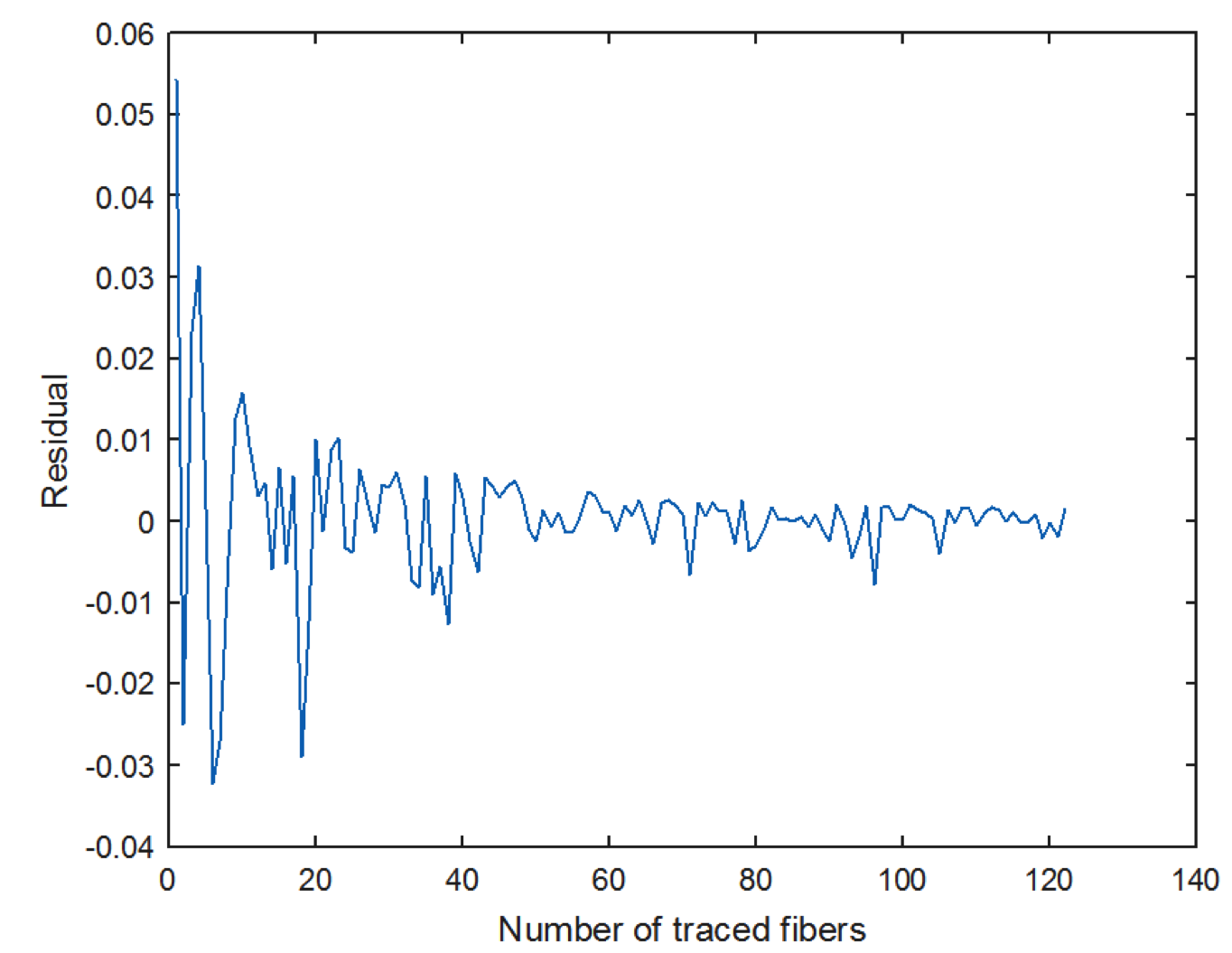}
\caption{Residual $e(j)$ as a function of number of fibers considered in calculating fiber straightness. Definition of residual is given in Eq. \ref{eq:residual}}
\label{fig:5}      
\end{figure}

Fiber arc length (s) was determined for each fiber tracing.  Cord length (L) was defined as the length of a best linear fit line to the same segment. Fiber straightness was defined as the ratio of chord length to arc length $L/s$ \cite{Hill2012}.
A fiber was designated as recruited to load bearing when its straightness reached 0.98 \cite{Hill2012,Hillthesis}. 
Careful attention was given to assuring the fiber straightness results for each sample at each load were independent of the number of fiber tracings. In particular, a mean value of straightness of the first $j$ tracings was defined as, 
\begin{equation}
m(j)  = \Frac{1}{j} \, \Sum{i=1}{j}   \Frac{L(i)}{s(i)}  \hspace*{.15in}  j = 1,2,3, \cdots n
\end{equation}
%
The difference between the calculated straightness for $j+1$ tracings and $j$ tracings was defined as a residual of $m(j)$:
\begin{equation}
e(j) = \Frac{m(j+1) -m(j)}{m(j+1)},  \hspace*{.15in}  j = 1,2,3, \cdots n-1
\label{eq:residual}
\end{equation}
When the residual satisfied $\mbox{abs}[e(j)] < 0.005$, the number of fiber tracings was deemed sufficient to evaluate fiber straightness. A representative plot for $e(j)$ is shown in Fig. \ref{fig:5} and illustrates how the bounds on the residual diminish with increasing number of fiber tracings, tending toward zero. 

\subsection{Quantification of the flattening of the lamina propria}  
The mucosa of the unloaded bladder wall is wavy, Fig. \ref{fig:1}, and gradually flattens under load.   
The flatness/waviness of the LP was assessed  in 3D reconstructions of the MPM images by calculating  the vertical position of the surface.  The lamina propria was defined to be flat when the standard deviation  of the vertical  position was less than 20 microns.
\begin{figure*}[!ht]
{\centerline{\includegraphics[width=6.25in]{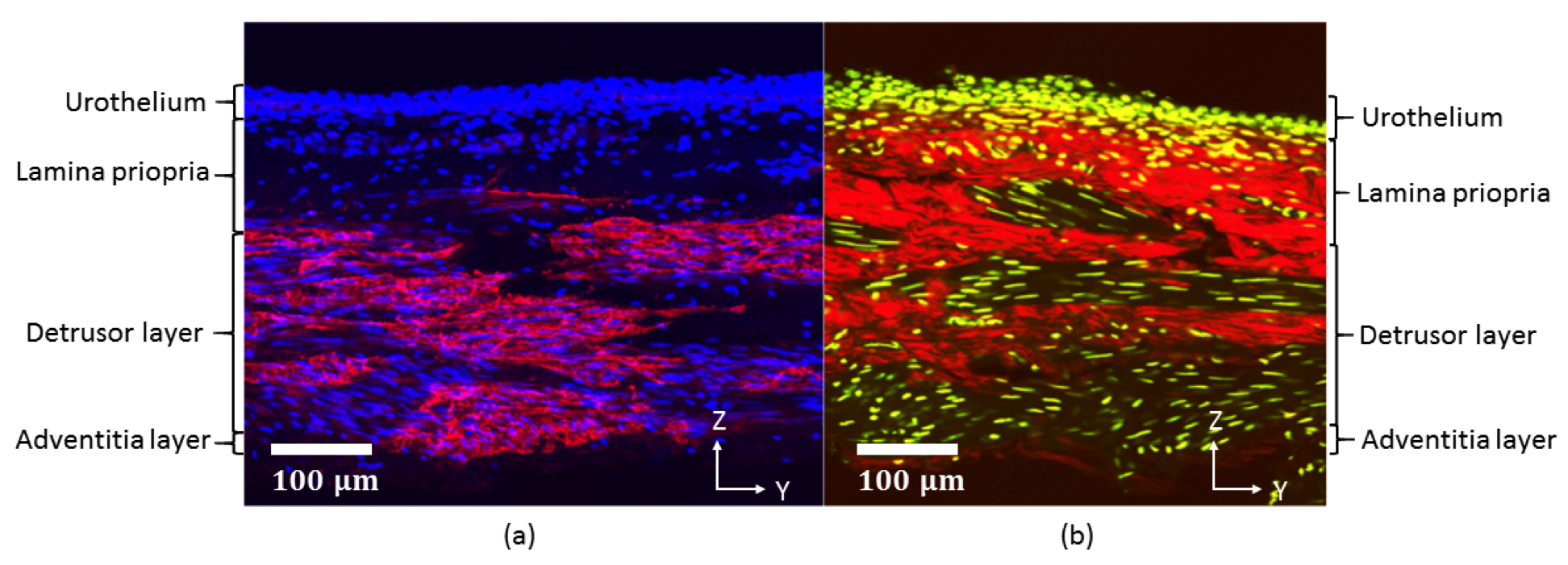}}}
\caption{Bladder cross section imagined using  (a) Confocal microscopy   and (b) Multiphoton Microscopy.  In (a), cell nuclei are shown in blue (DAPI), while red shows regions positive for $\alpha$SMA. Cell nuclei are yellow in (b) with collagen fiber signal in red.  The y-z plane corresponds to the longitudinal and transmural directions, respectively. Lumen side is on the top of the image.}
\label{fig:6}    
\end{figure*}
\subsection{Immunohistochemistry}
Immunofluorescent (IH) staining was performed to identify regions of smooth muscle cell bundles and location of cell nuclei, following established protocols from the  Center for Biologic Imaging (CBI),  (details can be found at   {www.cbi.pitt.edu/Protocols.htm}). Briefly, 
after preparing the square sample for mechanical testing, a portion of the remaining tissue was   fixed in 4\% PFA (paraformaldehyde) for 10 mins at room temperature. Cross sections were obtained by freezing the specimen in OCT and 2 \micronsp thick slices were cut using a cryostat (HM 50SE, Microm).
The sections were permeabilized with 0.2\% Triton-X 100 for 5 mins at room temperature,  washed three times with 1X PBS and incubated  in 5\% normal goat serum (NGS) overnight at 4$^o$C in a humidified chamber. 
The next day, the sample was incubated (diluted by 1\% NGS) with primary antibody  for $\alpha$SMA  (Sigma C6198) for 45 mins at 37$^o$C.  To stain cell nuclei, the slices were then washed with 1X PBS 3 times and incubated for 20 minutes without illumination  in DAPI (4$'$,6-diamidino-2-phenylindole, 1:1000, Life Technologies). All sections were then imaged using Olympus Fluoview 1000 confocal microscopy   (Olympus Imaging America, Melville, NY). A  neighboring section similarly stained with DAPI but without antibody for $\alpha$SMA was imaged using  multiphoton microscopy.
\section{Results}
\subsection{Overview of collagen fiber and cellular distribution across the wall layers}
\label{sec:2.1}
To provide context for discussions of the enface images of collagen fibers, a cross section of the rat bladder is provided in Fig. \ref{fig:6}, with IH staining imagined under confocal microscope and a neighboring section imaged using MPM.   Fig. \ref{fig:6}(a)  shows cell nuclei (blue) and $\alpha$SMA positive regions in red whereas Fig. \ref{fig:6}(b) shows the cell nuclei (yellow) and collagen fibers (red).  In Fig. \ref{fig:6}(a), moving from the lumen downwards, the cell density diminishes as one moves through the mucosa from the urothelium to the lamina propria.  The adjacent detrusor layer has large regions of $\alpha$SMA positive staining indicating areas of smooth muscle bundles.  Sparse cell nuclei can be seen in the adventitial layer.
 In Fig. \ref{fig:6}(b), consistent with Fig.\ref{fig:6}(a), the urothelium layer has a dense display of cell nuclei. Under MPM, the collagen within the lamina propria can be seen.  Moving into the detrusor layer, regions of dense collagen fibers are interspersed with regions populated with cells corresponding to regions of smooth cell bundles in Fig. \ref{fig:6}(a).  Further outwards, the collagen within a thin adventitial layer can be seen.  This sample has been gently flattened before fixation, so waviness  is not seen in the mucosa layer.
 \subsection{Mechanical response of the bladder wall}
 \label{sec:2.4new}
\begin{figure*}[!th]
{\centerline{\includegraphics[width=6.5in]{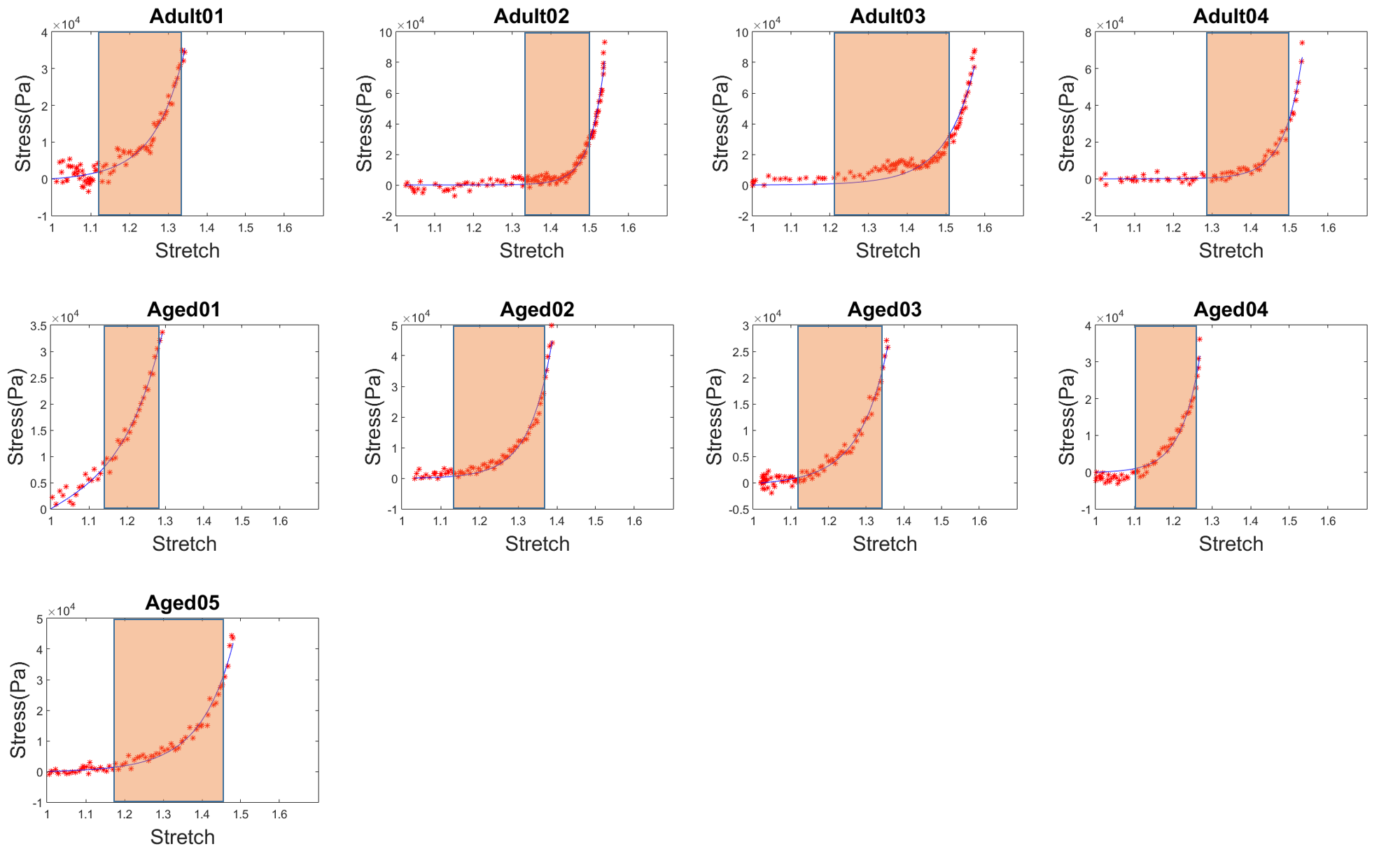}}}
\caption{Mechanical loading curves for bladder tissue showing Cauchy stress as a function of stretch ($\lambda$) with raw data (red) and exponential model from Eq. \ref{eq:exp} (blue).  Material parameters in Table \ref{table:properties}. The transition regime  for each curve is delineated by a colored rectangle. High stiffness regime following the transition regime inhibits further  extension. }
\label{fig:1x}    
\end{figure*}
The raw data for the biaxial testing for each sample are shown in Fig. \ref{fig:1x}, averaged for the longitudinal and circumferential directions.  This data was well fit by the exponential model  (\ref{eq:exp}) with an  R$^2 > 0.94$  for all samples, Table  \ref{table:properties}.
\begin{table}[h]
\footnotesize
\centering
\caption{Mechanical properties for each bladder sample.   Material constants $\mu,\gamma$ and corresponding R$^2$ for fit of constitutive model in Eq. \ref{eq:exp} to data shown in Fig. \ref{fig:1x}.   Strains $\epsilon_1$ and $\epsilon_2$ separate the low strain, transition and high strain  regimes of the loading curve.} 
\label{table:properties}
 \footnotesize
\begin{tabu}{l|ccc|cc|c}
 \footnotesize
 Sample  & $\mu$ (Pa)   & $\gamma$ & R$^2$ & $\epsilon_1$ & $\epsilon_2$ &Type\\  \toprule
 Adult01 & 4635 & 2.2     & 0.95                & 0.13         & 0.40         &II         \\
 Adult02 & 22   & 4.2     & 0.98                & 0.40         & 0.63           &I         \\
 Adult03 & 625   & 2.0     & 0.94                & 0.24         & 0.64            &I         \\
 Adult04 & 90    & 3.3     & 0.98                & 0.34         & 0.64            &I        \\
 Aged01  & 11,015 & 1.4     & 0.99                & 0.15         & 0.34           &II         \\
 Aged02  & 1814  & 2.9     & 0.98                & 0.16         & 0.44        &II         \\
 Aged03  & 4050  & 2.2     & 0.98                & 0.13         & 0.451         &II         \\
 Aged04  & 7720  & 3.7     & 0.96                & 0.11         & 0.31            &II         \\
 Aged05  & 1500  & 1.8    & 0.98                & 0.20         & 0.57            &I       
\end{tabu}
\end{table}

\subsection{Two bladder wall types identified based on wall extensibility.}
\begin{figure}[!ht]
\includegraphics[width=3.25in]{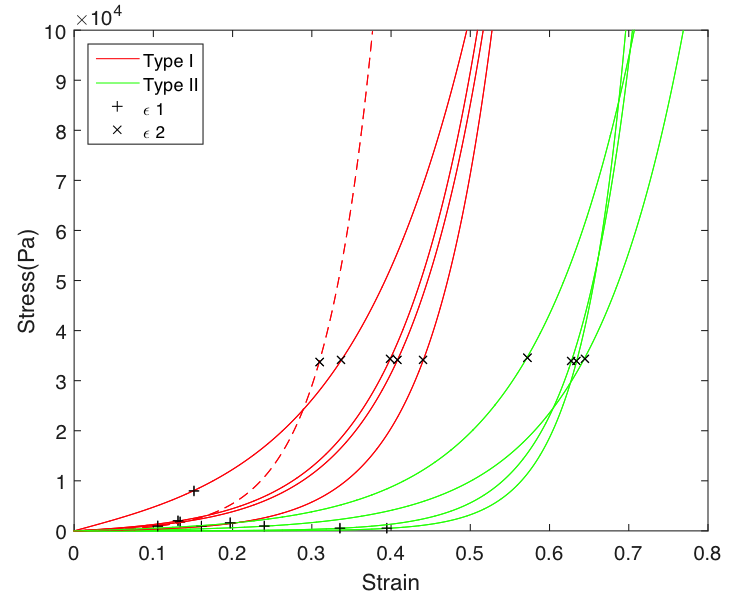}
\caption{Cauchy stress as a function of strain for  wall types I (green) and II (red) corresponding to composite of curves from Fig. \ref{fig:1x}.  The strain at the end  of toe regime ($\epsilon_1$)  and end of transition regime ($\epsilon_2$) are designated by ``+''  and ``x', respectively. } 
\label{fig:13}       
\end{figure}

The  strains $\epsilon_1$ and $\epsilon_2$ at the interface between the three loading regimes (toe, transition, and high-stress regimes) are given in Table \ref{table:properties}.  Two different categories of bladder were defined based on these strains.    In particular, the method of k means clustering  was used  to partition  the $n=9$ observations into $k=2$ clusters, where each observation belongs to the cluster with the nearest mean.  Type I has mean values ($\epsilon_1 =  0.29, \epsilon_2 = 0.62$), while \typeII has  means ($\epsilon_1 =  0.14, \epsilon_2 = 0.38$), Table \ref{table:properties}. A student T-test showed these groups were statistically distinct, (p=0.039 for $\epsilon_1$, and p=0.00001 for $\epsilon_2$).  Type I wall  is more extensible due to longer toe regimes (larger $\epsilon_1$) and larger strains before the onset of the high stress regime ($\epsilon_2$).   These differences can be seen visually for each bladder in Fig. \ref{fig:1x} and for the combined data set in Fig. \ref{fig:13}.  The blinded separation of the samples into two wall types is clearly seen in  Fig. \ref{fig:13}, where data for Type I are shown in green and for Type II in red.

\subsection{Bladder compliance can be maintained despite increased wall stiffness}
\begin{figure}[!ht]
\includegraphics[width=3.25in]{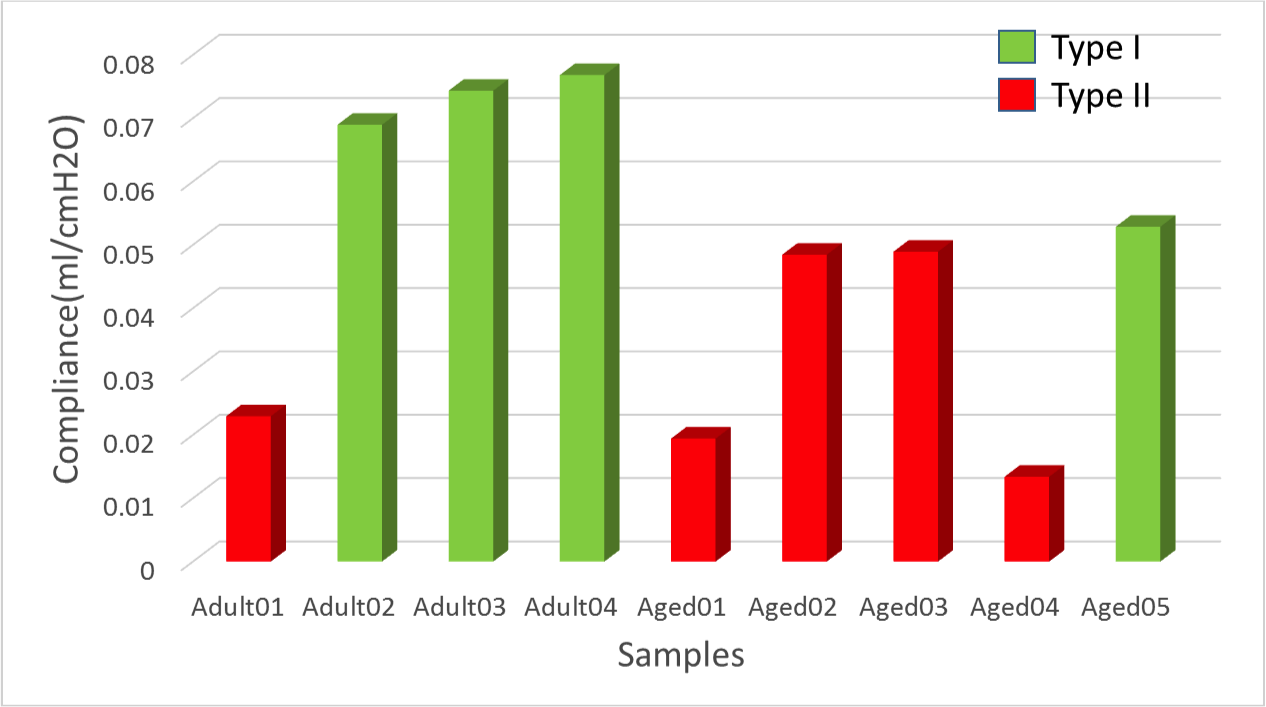}
\caption{Compliance of \typeI (green) and \typeII (red) bladder walls. }
\label{fig:compl}       
\end{figure}

  The bladder capacity, defined here as the bladder volume at cystometric pressure, is an important variable for the health of the animal since it determines how much urine can be stored comfortably before voiding.   As noted earlier, bladder compliance and capacity are proportional for a fixed cystometric pressure.   It follows from a simple equilibrium balance that
 reduced capacity can arise due to increased stiffness ($\mu, \gamma$), decreased volume ($V_o$) and/or  increased wall thickness $t_o$. 
  The average compliance (and hence capacity)  of the more elastic \typeI bladders is nearly twice that of  \typeII bladders  and these groups are statistically distinct (p=0.0038),  Table \ref{table:compl} and Fig. \ref{fig:compl}.  
\par
 Two of the (\typeII) bladders (Aged02 and  Aged03)  maintained a relatively large capacity (compliance) despite lower elasticity through increased $V_o$, Fig. \ref{fig:compl} and Table \ref{table:compl}.   The wall thickness was within one standard deviation of the average for \typeII walls and therefore did not play a large role in maintaining a high bladder capacity.  These two bladders had  the largest unloaded volumes of the entire cohort and $V_o$ was more than 30\%  greater than the  average volume of the \typeI bladders, Table \ref{table:compl}.   As a result, the  capacities of these \typeII bladder were within  70\% of the average for \typeI bladders. In contrast, the other three \typeII walls  were not enlarged relative to the \typeI bladders and their capacities were reduced to  less than 34\%   of the average capacity for the type I walls.  
 \par
\begin{table}[h]
\footnotesize
\centering
\caption{ Comparison of the unloaded wall thickness $t_o$, unloaded volume $V_o$, volume expansion V-$V_o$, and compliance C between the two wall types}
\label{table:compl}
 \footnotesize
\begin{tabu} to 0.44\textwidth {  m{1.15cm}  X[c]   X[c] X[c]  X[c]  X[c]  X[c]   }
 \footnotesize
	&  Wall & $t_o$ & $V_o$  &  $V-V_o$  &  C \\  
 Sample                &  Type  & (mm)   &(mm$^3$)   &{(mm$^3$})&   (ml/cmH$_2$O) \\ \toprule
Adult02  & I   & 0.52  & 170                               &414 &  0.069          \\
Adult03  & I  &0.61  & 166                              & 446 &  0.074           \\
Adult04  & I  &0.46    & 182                           & 460 &  0.077            \\ 
Aged05   & I & 0.66   & 143                            & 317 &  0.053       \\   \midrule
Average  & I  &   0.56  & 165 & 409 &          0.068    \\  \midrule
Std & I & 0.09& 17& 65&0.011 &\\  \midrule
Adult01  & II  & 0.66 &115                             & 138   &  0.023             \\  
Aged01   & II &  0.72 & 129                             & 116 &  0.019              \\
Aged02   & II  &  0.85   & 214                            & 291 & 0.048           \\
Aged03   & II  & 0.92  & 245                            & 293 &  0.049             \\
Aged04   & II  &1.12 & 151                             & 80 &  0.013             \\  \midrule
Average  & II    & 0.85 & 171 &  184&          0.031       \\ \midrule
Std &II &0.18 &56 &101 &0.017 &\\
 \bottomrule
\end{tabu}
\end{table}

\subsection{Relationship between  flattening of rugae and bladder loading curves}
\label{sec:2.2}
\begin{figure}[!ht]
\includegraphics[width=3.25in]{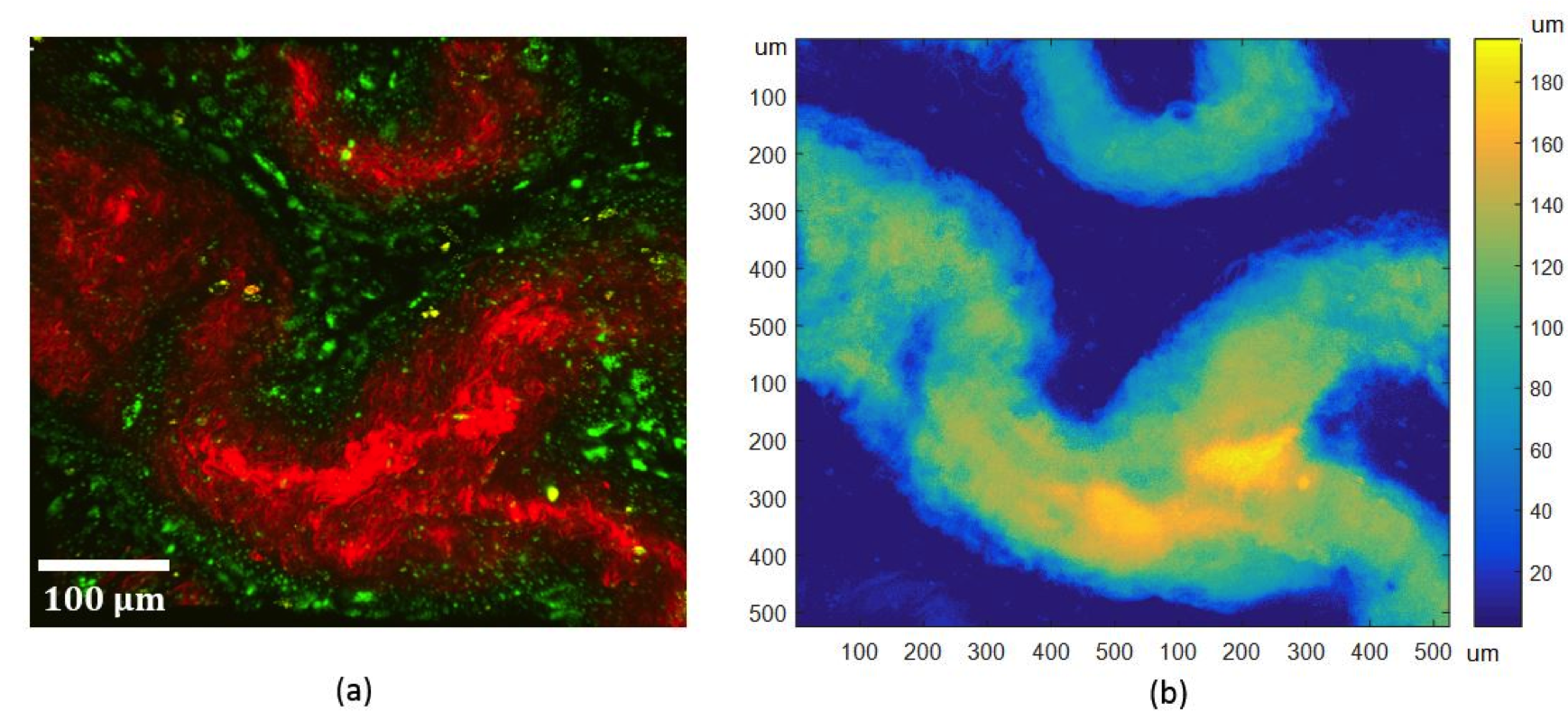}
\caption{Rugae of the unloaded bladder wall as seen in (a) Projected stack of multiphoton images with collagen fibers in red and  (b) Corresponding contour plot of surface elevation with peaks of  undulated surface (large z-value) shown in yellow.}
\label{fig:7}       
\end{figure}

A representative projected stack of MPM images as viewed from the lumen side of the bladder wall in the unloaded state is seen  Fig. \ref{fig:7}(a),
The  waviness of the unloaded luminal surface (rugae) can be seen with the undulated collagen fibers  in red.   
Since the depth of the stack is less than the depth of the unloaded LP layer, these undulations are truncated, leading to areas absent of the red signal in this projected view.  In Fig. \ref{fig:7}(b), the surface elevation of the luminal surface is shown as a contour plot for the 3D reconstructed stacks in Fig.\ref{fig:7}(a).  The surface folds are clearly multi-dimensional, not, for example, ridges running in the longitudinal direction.    
\subsubsection{Rugae are not flattened until transition regime}
\label{sec:2.4}
\begin{figure}
\includegraphics[width=3.25in]{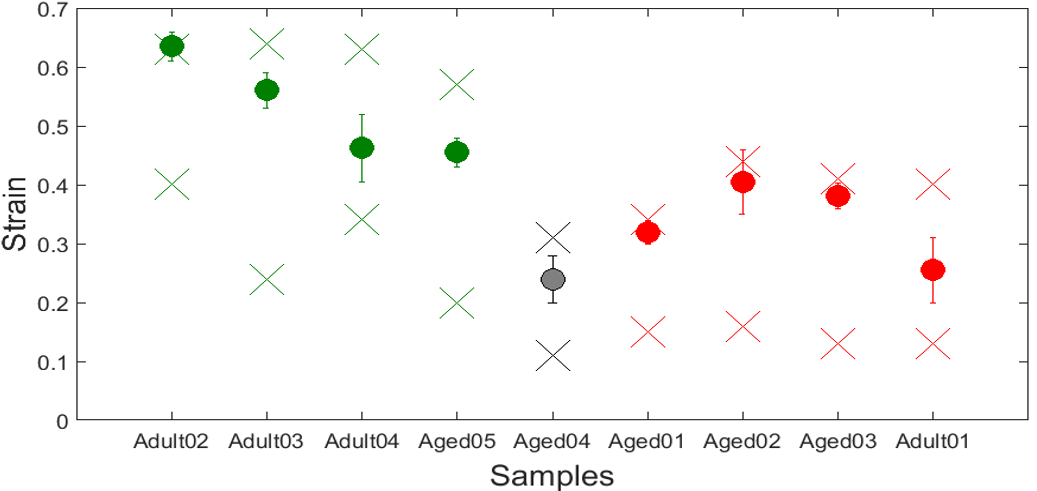}
\caption{Strain at which the undulations in the luminal surface (rugae) were lost for each of the nine cases.  The two ``x'' marks for each case  denote the strain ($y$-axis) at which the transition regime begins and ends.  The ``o'' denotes the strain at which the luminal surface was flattened.  Green, red and black correspond to wall types I, II, and \III, respectively.}
\label{fig:12}       
\end{figure}

The regime in the loading curve where rugae were lost for each sample is seen in 
Fig. \ref{fig:12}.  The two ``x'' marks denote the strain ($y$-axis) at which the transition regime begins and ends.  The ``o'' denotes the strain at which the luminal surface was flattened.  Since the imaging was done at discrete strain values, for each sample, there is a bar, signifying the largest strain at which the tissue was wavy and the smallest strain at which it was imaged as flat.  In all cases, the flattening occurred beyond the toe region,  within the transition regime, Fig. \ref{fig:12}.  We found that LPs of \typeII walls were flattened at lower strains in general than \typeI walls. 
\subsection{Collagen fiber recruitment across the bladder wall}
\begin{figure*}[!ht]
\includegraphics[width=6.25in]{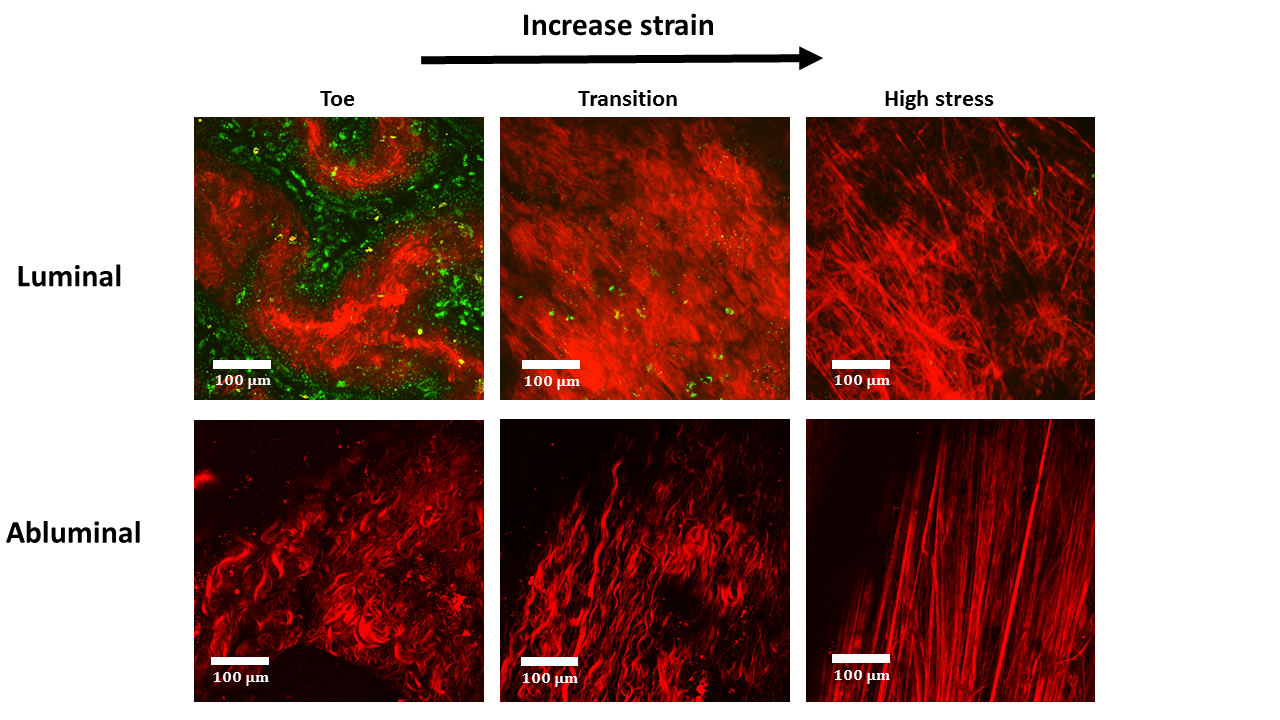}
\caption{Projected stacks of MPM images showing Adult04  as viewed from the luminal (row 1) and abluminal (row 2) sides.   Panels one to three correspond to the toe, transition and high stress regimes, respectively. With increasing strain,  undulations (rugae) of the mucosa on the luminal side are flattened (row 1), followed by straightening (recruitment) of the collagen fibers (column 3). 
}
\label{fig:8}       
\end{figure*}
Fig. \ref{fig:8} shows the fiber recruitment  with increasing strain  as seen in projected stacks of MPM images from both the luminal side (row 1) and abluminal side (row 2).  From the luminal side, both the flattening of the undulations  as well as the straightening of the collagen fibers in the LP and onset of recruitment can be seen.  From the abluminal side, highly coiled fibers can be seen in the DSM layer for the  toe regime with gradual straightening of the collagen fibers with increasing strain.  
\par

\subsubsection{LP and DSM collagen  recruitment is coordinated in elastic bladder walls}
\label{sec:2.3}
\begin{figure*}[!ht]
\centerline{\includegraphics[width=7.00in]{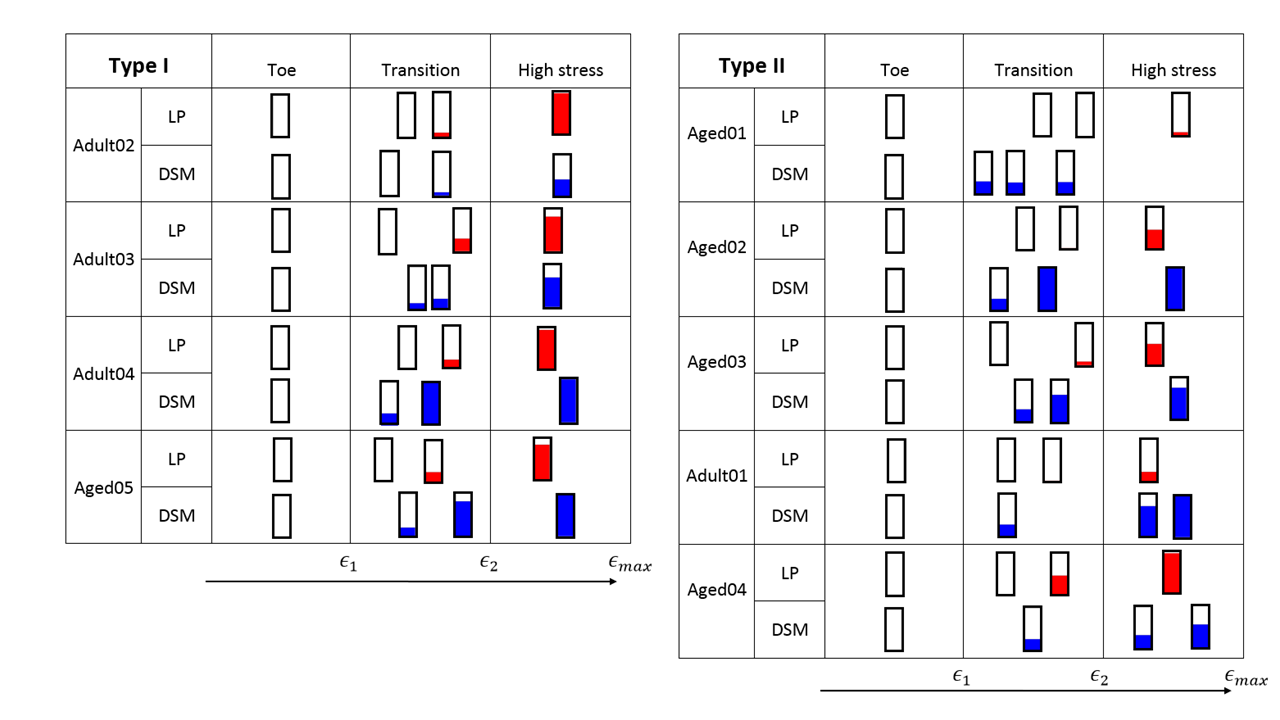}}
\caption{Fraction of recruited fibers at 4 different strain points in the LP and DSM layers for the toe, transition and high stress regimes of the loading curves for each case.   Here the fraction of recruited fibers is shown in red (LP layer) and blue (DSM).   Any empty box (e.g. toe regime) signifies that no fibers were recruited while a completely filled box signifies that all fibers were recruited.}
\label{fig:10}       
\end{figure*}

In order to understand the  relative role of collagen recruitment in the DSM and LP layers during loading, we quantified the percentage of recruited collagen fibers from both the luminal (lamina propria) and abluminal sides (detrusor layer) and considered these with respect to the toe, transition and high stress regimes, Fig. \ref{fig:10}.  
A representative fiber tracing for the abluminal side was shown in Fig. \ref{fig:4}. 
The fraction of recruited fibers at 4 different strain points in  each of the LP and DSM layers is shown in Fig. \ref{fig:10}. 
Since the MPM images were performed independently on luminal and abluminal sides, the data for the LP and DSM  layers are obtained at slightly shifted strains within each regime, and this is reflected in the horizontal location of the recruitment data within a regime.
\par
No fibers were recruited in the toe regime for any of the samples (n=9), Fig. \ref{fig:10}.  
In \typeI bladders, recruitment of collagen fibers was coordinated in the LP and DSM layers. In particular, collagen fibers from both the lamina propria and detrusor layer began to be recruited in the transition regime and most of the fibers were straightened in both layers in the high stress regime. For all \typeII samples except one (Aged04), the recruitment of collagen in the LP and DSM layers was uncoordinated.   Namely, few LP collagen fibers were recruited in the transition regime and the recruitment of DSM fibers dominated this regime.   The under recruitment of the DSM collagen continued through the high stress regime for these walls  In contrast, the \typeII bladder (Aged04)  showed a coordinated fiber recruitment  to \typeI bladders and is referred to as \typeIIp wall in further discussions and shown as dashed red lines in Fig. \ref{fig:10}. The \typeIIp wall (Aged04) had the earliest flattening of the  LP layer, Fig. \ref{fig:12},  and  the shortest toe regime with $\epsilon_1($\III$)= 0.11$, Table  \ref{table:properties}.
\par
\begin{figure}[!ht]
\includegraphics[width=3.25in]{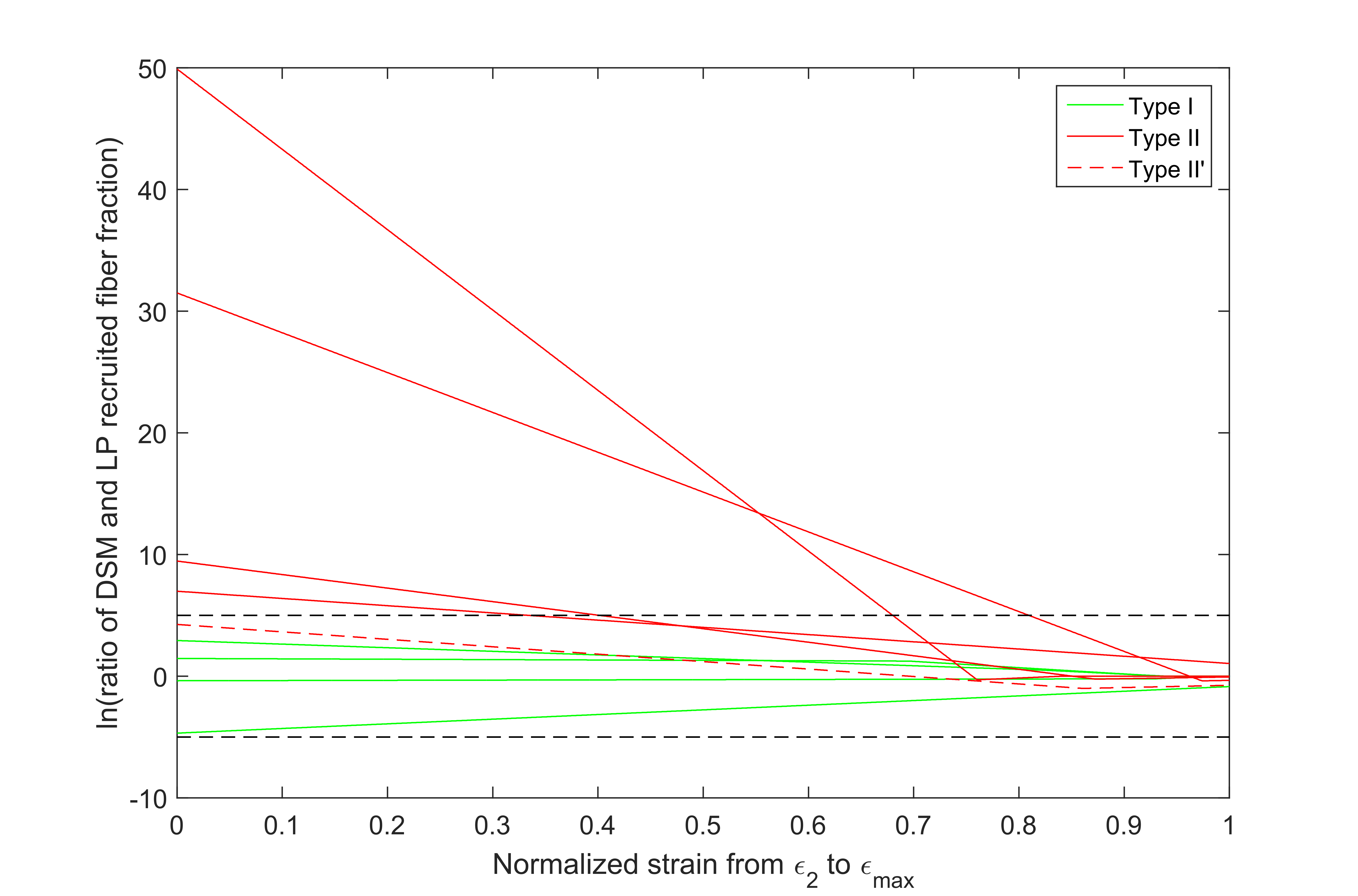}
\caption{Evaluation of the coordination of collagen recruitment between the DSM and LP layers.  The y-axis is the natural log of the ratio of fraction of recruited fibers in the DSM layer to those in the LP layer.  A value of zero would indicate perfect coordination.  For reference, the dashed black lines are plus and minus five on the natural log axis. Since there is no recruitment in the toe regime, the strain is normalized over the range $[\epsilon_1,\epsilon_{max}]$.}
\label{fig:11}       
\end{figure}

The coordination of collagen recruitment is further quantified In Fig. \ref{fig:11} where  the ratio of fraction of fiber recruitment in the DSM and LP layers is shown as a function of normalized tissue strain. 
On the  natural log scale used here, a value of zero on the vertical axis would indicate perfect coordination. For reference,  dashed lines are drawn at plus and minus five on this natural log vertical axis.  The horizontal strain axis is shifted to cover the range [$\epsilon_1$ to $\epsilon_{max}$], since there is no collagen recruitment in the toe regime ($\epsilon < \epsilon_1$).  Strain is recalculated  relative to the transition regime (shifted by $\epsilon_1$).
The recruitment for \typeI walls is relatively coordinated through the loading process,
while
\typeII walls show a much earlier recruitment of fibers in the DSM layer compared with the LP layer.  This would suggest the DSM layer is shielding the LP layer in these \typeII walls.
The  \typeIIp response  (red dashed line) in Fig. \ref{fig:11}  
fell within the coordinated recruitment region with the \typeI walls, consistent with results in Fig. \ref{fig:10}. 
\subsection{Aged versus adult  bladder}
\label{sec:2.4}
The bladders from the adult rats were more likely to be \typeI (3 out of 4), whereas the aged bladders were generally \typeII (4 out of 5). The aged bladders were thicker than the adult bladders (p=0.019) with averages of 0.89 $\pm$ 0.18 mm and 0.56 $\pm$ 0.09 mm.    The \typeI and \typeII walls had average thicknesses of  0.56 $\pm$ 0.09 mm  and 0.79$\pm$ 0.12 mm, respectively.   The type \IIIsp wall, an aged bladder, was substantially thicker than the other walls, with a thickness of 1.12 mm.   In addition, we found the wall thickness of adult samples are negatively correlated (R$^2$= -0.85) to their initial volume while the aged samples, except for the unusual case Aged04 (type II$'$),  showed the opposite  trend (R$^2$ = 0.95).
\section{Discussion} 
%
The bladder  is one of the most compliant organs in the body. For example,  a normal rat bladder can expand to about three times  its initial volume.  The same passive components of the bladder wall, such as elastin and collagen fibers,  are also found in  the walls of other biological tissues.  It is the wall architecture, or the way these components are combined, that must be responsible for its extensibility.  In this work, we have made use of a custom designed planar biaxial system compatible with multiphoton microscopy to directly investigate the relationship between bladder wall structure and wall elasticity during loading in fresh bladder samples from adult and aged rats.
\par
To understand the changing roles of the wall layers during loading, we defined three regimes for the bladder mechanical loading curves: the  toe regime, transition regime, and high stress regime. 
Since the  bladder material stiffens increases so rapidly in the high stress regime 
the bladder extensibility  is largely determined by the length of  the toe  and transition regimes, Fig. \ref{fig:13}.  
Namely, the  steep increase in wall stiffness in the high stress regime  inhibits further substantial bladder expansion, essentially
``locking in'' the maximum bladder volume.
\smallskip \\  \noindent {\bf Two wall types- classified by extensibility}  \\
Within 5 aged samples and 4 adult samples, we found two types of bladders,  defined by  measures of their extensibility,  $\epsilon_1$ and $\epsilon_2$.   
 Type 1 walls  have longer toe regions and later onset of the high stress regime compared with Type II and these differences reached statistical significance. 
The increased elasticity of \typeI walls is mainly due to  significantly longer toe regimes than \typeII bladders.  The average length of the transition regime is also longer, but it does not reach statistical significance.  
\smallskip \\   \noindent {\bf Mechanisms for high elasticity in the toe regime} \\
The high elasticity of the toe regime is only possible if  both the inner (mucosa) and outer (DSM and adventitial) layers of the wall are highly elastic.   
 In this work, we  showed that the high elasticity of the  LP layer is due to folds (rugae) in the bladder wall.  With increasing strain the LP gradually flattens with little resistance to load.  In parallel, the LP collagen fibers become more planar and eventually begin to straighten.  Importantly,  the rugae do not fully flatten, nor does collagen recruitment commence until after the end of the toe regime (in the transition regime). 
 Therefore,  the bladder wall has two levels of undulations to provide extensibility: wall undulations (rugae) on the order of 100 microns, Fig. \ref{fig:7}, as well as undulations in the collagen fibers themselves, on the order of tens of microns.    This second level does not engage until the after the first (rugae) are flattened.  
%
%
\par
Previous work has suggested  the flexibility of the outer wall  is provided by the architecture of the SMC bundles and collagen fibers. In particular, that in the unloaded bladder the SMC bundles are connected by wavy collagen fibers, \cite{Chang1999}.  Hence, the   reorientation and stretching of the SMC bundles as well as straightening of the  wavy collagen fibers are mechanisms that provide high flexibility in the DSM layer.
  Our results are consistent with  this conjecture.  In particular, while we did not investigate the connection between the SMC bundles and collagen fibers in this work, we have shown the DSM collagen fibers were not recruited until the transition regime was reached. This is consistent with  the proposed mechanism for large extensibility of this layer.
   A schematic of  mechanisms for extensibility of the LP and DSM layers is shown in Fig. \ref{fig:14}.
\smallskip \\   \noindent {\bf Mechanisms responsible for stiffening in the transition regime of \typeI bladders} \smallskip \\
 In the transition regime of \typeI walls, the stress increase is due to the gradual fiber recruitment of fibers in both layers as demonstrated in Fig. \ref{fig:11}.   The fiber recruitment in the LP and DSM layers are coordinated in that both layers demonstrate collagen recruitment that continues to substantial levels in the high stress regime, Fig. \ref{fig:10}.  
%
\smallskip \\  \noindent {\bf Physical mechanisms responsible for loss of bladder extensibility (\typeII and \typeIIp walls)} \smallskip \\
The central difference in the \typeII and \typeIIp walls compared with the \typeI walls was the shortened toe regime. 
Since the toe regime is a region of high extensibility under low loads, 
even a small amount of loading of stiff collagen fibers will  end this regime and force it to enter the transition regime. 
We found that \typeII bladders have a shorter toe regime because the DSM fibers were straightened prematurely (lower strains than \typeI walls), ending the toe regime, Figs. \ref{fig:10} and \ref{fig:11}.   In contrast, in the \typeIIp wall, fibers from both the LP and DSM layers were  recruited prematurely, resulting in the earliest flattening of the LP among all the samples, Fig. \ref{fig:12}. As  result, this case had the shortest toe regime.
\par
\begin{figure*}[!ht]
	\centerline{\includegraphics[width=5.00in]{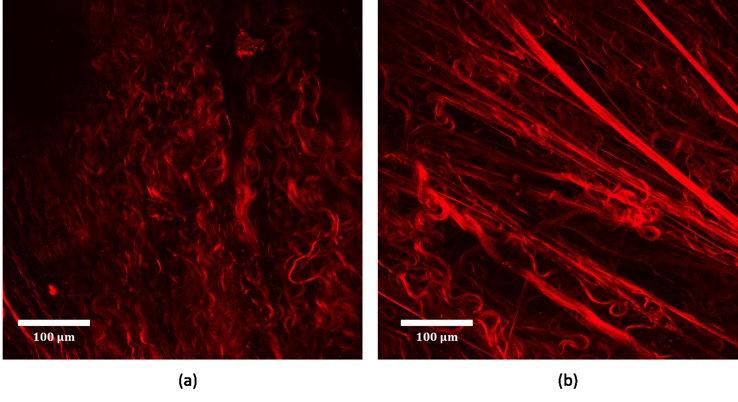}}
	\caption{Collagen fibers in  \typeII wall (Aged01) as seen in projected stacks of  abluminal MPM images  (a) toe regime and (b) transition regime.  Unbalanced fiber recruitment is seen in the  DSM where  some fibers are fully straightened and others are highly tortuous.}
	\label{fig:15}       
\end{figure*}
\begin{figure*}[!ht]
	\centerline{\includegraphics[width=6.50in]{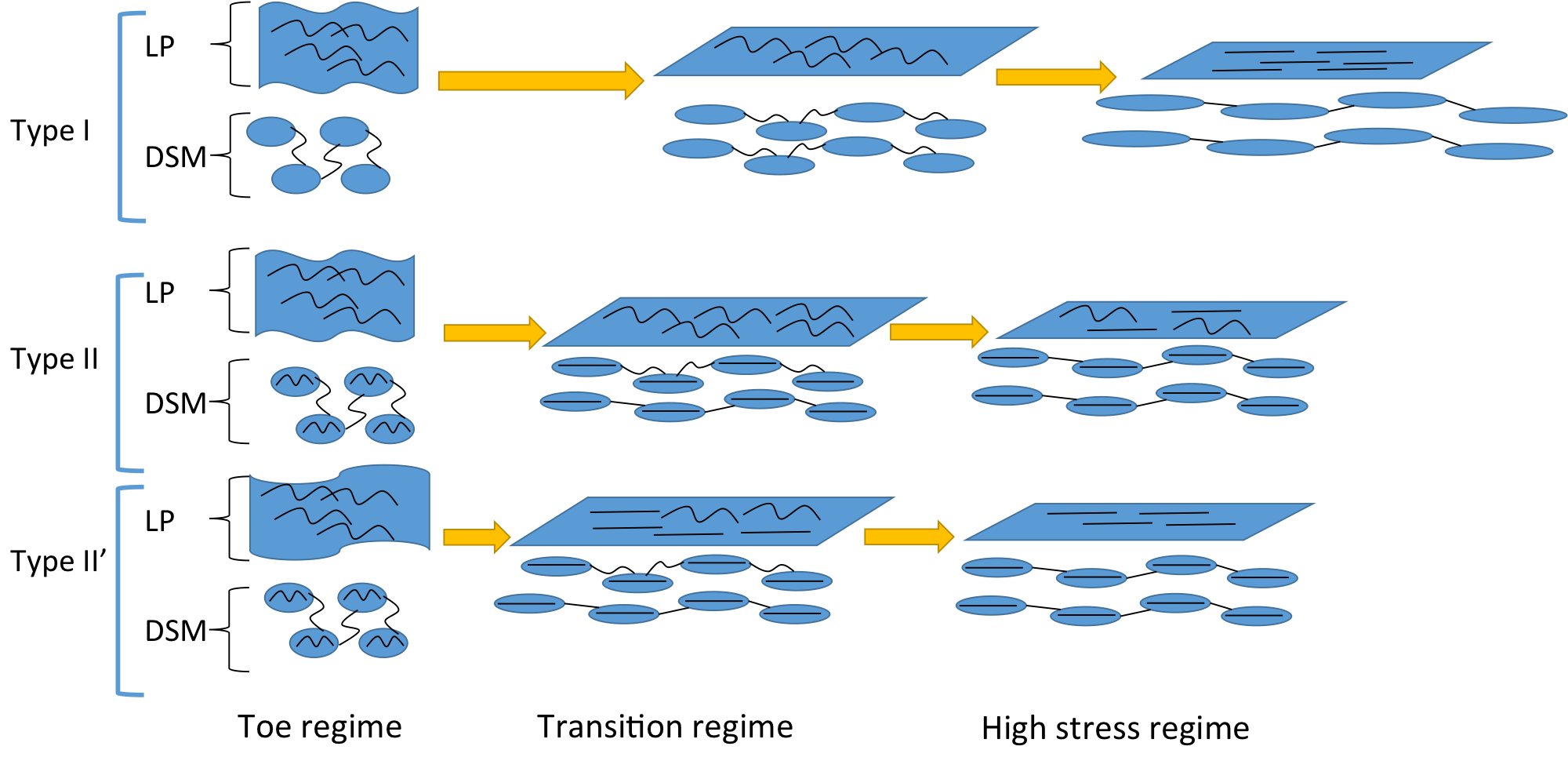}}
	\caption{Schematic of the  findings for the relative roles of the collagen fibers in the LP and DSM layers by wall type.  In all wall types,  the toe regime  is characterized by flattening of the rugae without fiber recruitment.  In the toe regime, the extensibility of the DSM is conjectured to arise from the wavy nature of the collagen fibers connecting flexibly SMC bundles. For the transition regime of  \typeI walls, the fiber recruitment in the LP and DSM layers are coordinated.  In \typeII walls, the toe region is cut short by early recruitment of the DSM collagen which  is conjectured to arise from infiltration of fibers in the SMC bundles as well as other stiffening mechanisms.  As a result, complete recruitment of LP collagen is inhibited, even in the high stress regime. In \typeIIp walls, although the LP and DSM fiber recruitment is coordinated,  the rugae are reduced,  resulting in the earliest recruitment.}
	\label{fig:14}       
\end{figure*}

In \typeII walls the early recruitment of DSM collagen fibers  prevented the LP fibers from engaging and straightening. In fact, large fractions of LP fibers remained unrecruited even in the high stress regime,  Fig. \ref{fig:10}.    In one sample (Aged01), we also saw a lack of coordination of fiber recruitment even within the DSM layer itself with some highly recruited 
fibers preventing continuing recruitment of other still highly tortuous fibers, Fig. \ref{fig:15}.  
This is consistent with the unchanging fraction of recruited DSM fibers over the transition regime in that sample, Fig. \ref{fig:11}.
\par
These findings  suggest  the DSM  layer of \typeII walls somehow lost one or more of the mechanisms for  structural transformation to accommodate extension before fibers are recruited. Mure and Galdo found an increased ratio of Type III to Type I collagen in non-compliant bladders \cite{Mure2004}. 
Chang found that in normal bladder, \typeIIp  collagen localization is largely confined to the 
 interfascicular regions (between SMC bundles) of the detrusor while in non-compliant bladder, \typeIIp collagen is found not only in the intrafasciular region, but also within the smooth muscle bundles  \cite{Chang1999}. The abnormal \typeIIp collagen fibers  in the SMC bundles may be the cause of the earlier fiber recruitment because they could stiffen the SMC bundles, forcing earlier recruitment of the interfascicular fibers. 
 \par
$\;$ \smallskip \\   \noindent {\bf Overview of structure/function relationship}  \\
The  relative roles of the collagen fibers in the LP and DSM layers, including conjectures regarding the role of the connection between SMC bundles and collagen fibers are shown in a schematic in Fig. \ref{fig:14}.
In the toe regime of all wall types, the LP is wavy and no fibers are recruited to load bearing.  The DSM layers collagen fibers are drawn to shown an interconnection with smooth muscle bundles (following the conjecture of \\ Chang \cite{Chang1998} and recent work on murine bladder using MPM \cite{Hornsby2016}). As shown in the present  work, during further loading of the \typeI wall,  the LP layer becomes flattened and  collagen fibers in both layers are gradually recruited to load bearing.   For \typeII walls,  the toe regime ends earlier than for \typeI walls and the  fiber loading is unsynchronized.  For \typeII walls,  some of the DSM collagen fibers become load bearing prematurely inhibiting the strains needed for the  collagen in the LP layer to become load bearing.   
$\;$ \smallskip \\   \noindent {\bf Effect of Aging}  \\
The heterogeneous aging process seen across species (with some animals aging more rapidly than others, for example, or in different ways), can confound analysis.   For this reason, we looked for wall types based on mechanical response, rather than trends with aging. 
Despite the small numbers of samples in this study, a high statistical significance was found,  demonstrating a propensity for aged walls to be Type II  and of increased thickness.
$\;$ \smallskip \\   \noindent {\bf Elasticity, bladder capacity and  bladder compliance}  \\
In this work, we have emphasized the need to consider   bladder elasticity separately from bladder capacity and bladder compliance. Like other soft tissues such as arteries,  the relationship between local stress  and strain is highly nonlinear and cannot be captured with a single measure of stiffness.   This is particularly important for the bladder, since  the the bladder empties and fills during normal filling/voiding cycles.   This is in contrast to, for example, an artery wall where the cyclic loading takes place about a largely inflated state.  
\par
 Bladder compliance, which is often used in the literature on the biology of the bladder wall,  is a  linear approximation to the bladder pressure/volume relationship.  
In particular, it is defined as the slope of in vivo bladder volume versus pressure curves, from the unloaded bladder to the bladder capacity (near the leak point). This is a very rough approximation to the mechanical response of the entire bladder.  It does not separate out the three regimes of loading, nor consider the wall properties independent of the response of the bladder as a whole.  While compliance may be the most amenable measurement in vivo, there are some concerns when it is used ex vivo for assessment of the bladder wall.  First, we have shown that the  bladder capacity is dominated by the {\em length} the toe regime, (not even the stiffness in this regime).    Secondly, we have found  a stiffer bladder  can  compensated  for  low elasticity through an increase in bladder size. Namely,  two bladders can have roughly the same compliance, yet one can have a shortened toe region  due to uncoordinated recruitment of collagen in the LP and DSM layers.
$\;$ \smallskip \\   \noindent {\bf Limitations and future directions}  \\
In this study,  we focused attention on a population of male Fischer rats.  In future studies, it will be important to extend this study to female rats as well as to human bladders.
While, we would have liked to have increased the population size for this study even further,  we were limited by the challenges associated with obtaining older rats as well as the time consuming aspects of the MPM analysis.   Nonetheless,  statistical significance was achieved relating wall type to structure of the extracellular matrix during loading.    
\par
Necessarily, this study was carried out ex-vivo.  There is currently no method for measuring collagen recruitment across the wall layers in vivo.   In the future, studies could be performed to evaluate recruitment in fixed, inflated bladders to provide further evidence to support the current findings. However, such studies would be limited by the need to consider different bladders for each loading level, rather than a single sample across all loading levels, such as was done in the present work.  Future work could also explore the structural mechanisms behind  bladder viscoelasticity and anisotropy \cite{Nagatomi2004,Parekh2010}.
\section{Conclusions}
This work has taken advantage of advances  in bioimaging  to  directly  assess collagen fibers and  wall architecture during loading.  The  large extensibility of the bladder was shown to be determined largely by the length of the toe region of the loading curve.
which arises from gradual flattening of the folds in the bladder  wall.  
These rugae   provide a mechanism for low resistance  flattening without any discernible recruitment of collagen fibers throughout the toe regime.  We have shown that, in contrast to prior conjectures on the bladder wall, that a coordinated recruitment of collagen across the LP and DSM layers is essential for the large elasticity of the bladder wall.   
Furthermore,   wall extensibility 
can  be lost by 
 premature recruitment of collagen in the DSM that cuts short the toe region.
Knowledge of the mechanisms responsible for bladder compliance is essential for providing targets for improved diagnostics,  developing   novel medical treatments of bladder dysfunction, and for understanding  the biomechanical environment of the intramural cells that drive changes in the bladder wall in health and disease.
 \section{Acknowledgements} 
The authors gratefully acknowledge the  NIH National Institute of Aging  for funding through 1R56 AG050408-01(PI Birder) as well as the aged rats used in this study.   The authors also gratefully acknowledge Mr. Chih Yuan Chuang for performing the immunohistochemistry work to obtain Figure 5.
\clearpage
\clearpage
 \bibliographystyle{spphys} 
 \bibliography{BladderBib4_21} 
\end{document}